\documentclass[a4paper,11pt]{article}
\usepackage{pos}

\title{Doubly charmed pentaquark states with strangeness $S=0, -1$}

\author*[a,b]{Wei Chen}
\author[a]{Feng-Bo Duan}
\author[c,d]{Zi-Yan Yang}
\author[e,f]{Qi-Nan Wang}
\author[a]{Xu-Liang Chen}
\author[c,d,b,g]{Qian Wang}

\affiliation[a]{School of Physics, Sun Yat-sen University, Guangzhou 510275, China}
\affiliation[b]{Southern Center for Nuclear-Science Theory (SCNT), Institute of Modern Physics, Chinese Academy of Sciences, Huizhou 516000, Guangdong Province, China}
\affiliation[c]{State Key Laboratory of Nuclear Physics and Technology, Institute of Quantum Matter, South China Normal University, Guangzhou 510006, China}
\affiliation[d]{Guangdong Basic Research Center of Excellence for Structure and Fundamental Interactions of Matter, Guangdong Provincial Key Laboratory of Nuclear Science, Guangzhou 510006, China}
\affiliation[e]{College of Physical Science and Technology, Bohai University, Jinzhou 121013, China}
\affiliation[f]{Editorial Department of Journal, Bohai University, Jinzhou 121013, China}
\affiliation[g]{Research Center for Nuclear Physics (RCNP), Osaka University, Ibaraki 567-0047, Japan}

\emailAdd{chenwei29@mail.sysu.edu.cn}

\abstract{In this work, we have studied the mass spectra of doubly charmed pentaquark states with strangeness $S=0, -1$ by using the method of QCD sum rules. We use the parity projected sum rules to separate the contributions of negative and positive parities from the two-point correlation functions induced by the pentaquark interpolating currents. Our results predict the existence of some potential doubly charmed pentaquark bound states.}

\FullConference{The XVIth Quark Confinement and the Hadron Spectrum Conference (QCHSC24)\\
 19-24 August, 2024\\
 Cairns Convention Centre, Cairns, Queensland, Australia\\}

\begin{document}
\maketitle

\section{Introduction}
In the past ten years, some candidates of hidden-charm pentaquark states have been observed by the LHCb collaboration. 
In 2015, LHCb reported $P_{c}(4380)$ and $P_{c}(4450)$ in the $J/\psi p$ invariant mass spectrum~\cite{Aaij2015}. Four years later, they further claimed that $P_{c}(4450)$ can be separated into two narrow structures $P_{c}(4440)$ and $P_{c}(4457)$ with the significance of $5.4\sigma$, while a new pentaquark state $P_{c}(4312)$ was reported at the same time~\cite{Aaij2019}. In the $B_{s}^0\to J/\psi \bar{p}p$ decays, an evidence for $P_{c}(4337)$ was found in both the $J/\psi p$ and $J/\psi \bar{p}$ final states~\cite{Aaij2022}. Recently, the $P_{cs}(4459)$ and $P_{cs}(4338)$ were reported as the candidates of hidden-charm pentaquark states with strangeness in the $J/\psi \Lambda$ invariant mass distribution in the $\Xi_{b}^-\to J/\psi K^- \Lambda$~\cite{Aaij2021} and $B^-\to J/\psi \Lambda \bar{p}$~\cite{LHCb:2022ogu} decay processes, respectively. On the other hand, LHCb also observed the doubly charmed tetraquark state $T_{cc}^+$ in the $D^0D^0\pi^+$ invariant mass spectrum~\cite{Aaij2022a,Aaij2022b}. 

These observations have extensively attracted research interest for tetraquark and pentaquark states in both hidden-charm and doubly charmed sectors~\cite{Chen:2016qju,Guo:2017jvc,Liu:2019zoy,Brambilla:2019esw,Guo:2019twa,Chen:2022asf,Liu:2024uxn,Zhu:2024swp}. To date, there have been some theoretical investigations on the existence of doubly charmed pentaquark molecular states in the charmed baryon-meson $\Lambda_c/\Sigma_c^{(*)}/\Xi_{c}^{('*)}D^{(*)}$ systems~\cite{Chen2017,Liu2020,Chen2021b,Chen2021a,Dong2021a,Shen2023,Shimizu:2017xrg,Duan:2024uuf,Yang:2024okq,Liu:2023clr,Yang:2020twg} and doubly charmed baryon-meson $\Xi_{cc}^{(*)}\pi/\eta/\rho/\omega$, $\Xi_{cc}\bar K$, $\Omega_{cc} K$ systems~\cite{Liu:2023clr,Yang:2020twg,Yang:2024okq,Guo:2017vcf}. 
 Besides, the doubly charmed pentaquarks have been also studied in the compact configurations of diquark-diquark-antiquark model~\cite{Zhou:2018bkn,Wang:2018lhz,Ozdem:2022vip} and triquark-diquark model~\cite{Xing:2021yid}. In this talk, I will introduce our recent studies of the doubly charmed pentaquarks with strangeness $S=0, -1$ by using the QCD sum rule method. 

\section{Formalism}
We construct the following interpolating currents to investigate the $\Lambda _{c}^{(*)}D^{(*)}$ and $\Sigma _{c}^{(*)}D^{(*)}$ molecular pentaquarks with strangeness $S=0$
\begin{equation}\label{currents1}
    \begin{split}
      &J^{\Lambda  _{c}D}=\varepsilon ^{abc}\left[\left(u_{a}^{T}\mathcal{C} \gamma _{\mu }c_{b} \right)\gamma _{5}\gamma ^{\mu }d_{c}-\left(d_{a}^{T}\mathcal{C} \gamma _{\mu }c_{b} \right)\gamma _{5}\gamma ^{\mu }u_{c}\right] \left[\bar{d}_{d}i\gamma _{5}c_{d}\right] \, ,\\
      &J_{\mu }^{\Lambda  _{c}D^*}=\varepsilon ^{abc}\left[\left(u_{a}^{T}\mathcal{C} \gamma _{\nu }c_{b} \right)\gamma _{5}\gamma ^{\nu }d_{c}-\left(d_{a}^{T}\mathcal{C} \gamma _{\nu }c_{b} \right)\gamma _{5}\gamma ^{\nu }u_{c}\right] \left[\bar{d}_{d}\gamma_{\mu }c_{d}\right] \, ,\\
      &J_{\mu }^{\Lambda  _{c}^*D}=\varepsilon ^{abc}\left[\left(u_{a}^{T}\mathcal{C} \gamma _{\mu }c_{b} \right)d_{c}-\left(u_{a}^{T}\mathcal{C} \gamma _{\mu }d_{b} \right)c_{c}\right] \left[\bar{d}_{d}i\gamma _{5}c_{d}\right] \, ,\\
      &J_{\mu \nu }^{\Lambda  _{c}^*D^*}=\varepsilon ^{abc}\left[\left(u_{a}^{T}\mathcal{C} \gamma _{\nu }c_{b} \right)d_{c}-\left(u_{a}^{T}\mathcal{C} \gamma _{\nu }d_{b} \right)c_{c}\right] \left[\bar{d}_{d}\gamma_{\mu }c_{d}\right]+\left(\mu \leftrightarrow \nu \right) \, ,\\
      &J^{\varSigma  _{c}D}=\varepsilon ^{abc}\left[\left(u_{a}^{T}\mathcal{C} \gamma _{\mu }c_{b} \right)\gamma _{5}\gamma ^{\mu }d_{c}+\left(d_{a}^{T}\mathcal{C} \gamma _{\mu }c_{b} \right)\gamma _{5}\gamma ^{\mu }u_{c}\right] \left[\bar{d}_{d}i\gamma _{5}c_{d}\right] \, ,\\
      &J_{\mu }^{\varSigma  _{c}D^*}=\varepsilon ^{abc}\left[\left(u_{a}^{T}\mathcal{C} \gamma _{\nu }c_{b} \right)\gamma _{5}\gamma ^{\nu }d_{c}+\left(d_{a}^{T}\mathcal{C} \gamma _{\nu }c_{b} \right)\gamma _{5}\gamma ^{\nu }u_{c}\right] \left[\bar{d}_{d}\gamma_ {\mu }c_{d}\right] \, ,\\
      &J_{\mu }^{\varSigma  _{c}^*D}=\varepsilon ^{abc}\left[2\left(u_{a}^{T}\mathcal{C} \gamma _{\mu }c_{b} \right)u_{c}+\left(u_{a}^{T}\mathcal{C} \gamma _{\mu }u_{b} \right)c_{c}\right] \left[\bar{d}_{d}i\gamma _{5}c_{d}\right] \, ,\\
      &J_{\mu \nu }^{\varSigma  _{c}^*D^*}=\varepsilon ^{abc}\left[2\left(u_{a}^{T}\mathcal{C} \gamma _{\nu }c_{b} \right)u_{c}+\left(u_{a}^{T}\mathcal{C} \gamma _{\nu }u_{b} \right)c_{c}\right] \left[\bar{d}_{d}\gamma_ {\mu }c_{d}\right]+\left(\mu \leftrightarrow \nu \right) \, ,
          \end{split}
\end{equation}
and the $\Xi_{c}^{(\prime\ast)}D^{(\ast)}$ molecular pentaquarks with  strangeness $S=-1$
\allowdisplaybreaks{
\begin{eqnarray}\label{currents2}
\nonumber \eta^{\Xi_{c}D}&=&\frac{1}{\sqrt{2}}\epsilon_{abc}\left[\left(u_a^TC\gamma_5s_b-s_a^TC\gamma_5u_b\right)c_c\right]\left[\bar{d}_d\gamma_5c_d\right],\\
\nonumber \eta^{\Xi_{c}^\prime D}&=&\frac{1}{\sqrt{2}}\epsilon_{abc}\left[\left(u_a^TC\gamma_\mu\gamma_5s_b-s_a^TC\gamma_\mu\gamma_5u_b\right)\gamma_\mu c_c\right]\left[\bar{d}_d\gamma_5c_d\right],\\
\nonumber \eta^{\Xi_{c}D^\ast}&=&\frac{1}{\sqrt{2}}\epsilon_{abc}\left[\left(u_a^TC\gamma_5s_b-s_a^TC\gamma_5u_b\right)\gamma_\mu c_c\right]\left[\bar{d}_d\gamma_\mu c_d\right],\\
\eta_{\mu}^{\Xi_{c}^\prime D^\ast}&=&\frac{1}{\sqrt{2}}\epsilon_{abc}\left[\left(u_a^TC\gamma_\nu\gamma_5s_b-s_a^TC\gamma_\nu\gamma_5u_b\right)\gamma_\nu c_c\right]\left[\bar{d}_d\gamma_\mu c_d\right],\\
\nonumber \eta_{\mu}^{\Xi_{c}^\ast D}&=&\sqrt{\frac{2}{3}}\epsilon_{abc}\left[(s^T_aC\gamma_\mu u_b)\gamma_5c_c+(u^T_aC\gamma_\mu c_b)\gamma_5s_c+(c^T_aC\gamma_\mu s_b)\gamma_5u_c\right]\left[\bar{d}_d\gamma_5c_d\right],\\
\nonumber \eta^{\Xi_{c}^\ast D^\ast}&=&\sqrt{\frac{2}{3}}\epsilon_{abc}\left[(s^T_aC\gamma_\mu u_b)\gamma_5c_c+(u^T_aC\gamma_\mu c_b)\gamma_5s_c+(c^T_aC\gamma_\mu s_b)\gamma_5u_c\right]\left[\bar{d}_d\gamma_\mu c_d\right],\\
\nonumber \eta_{\mu\nu}^{\Xi_{c}^\ast D^\ast}&=&\sqrt{\frac{2}{3}}\epsilon_{abc}\left[(s^T_aC\gamma_\mu u_b)\gamma_5c_c+(u^T_aC\gamma_\mu c_b)\gamma_5 s_c+(c^T_aC\gamma_\mu s_b)\gamma_5 u_c\right]\left[\bar{d}_d\gamma_\nu c_d\right]+(\mu\leftrightarrow\nu). 
\end{eqnarray}}
Besides, we also study the $\Xi_{cc}^{(\ast)}\bar{K}^{(\ast)}$ molecular states with strangeness $S=-1$ by using the interpolating currents
\allowdisplaybreaks{
\begin{eqnarray}\label{currents3}
\nonumber \xi^{\Xi_{cc}\bar{K}}&=&\left[\epsilon_{abc}(c^T_aC\gamma_\mu c_b)\gamma_\mu\gamma_5u_c\right]\left[\bar{d}_d\gamma_5s_d\right],\\
\nonumber \xi_{\mu}^{\Xi_{cc}\bar{K}^\ast}&=&\left[\epsilon_{abc}(c^T_aC\gamma_\nu c_b)\gamma_\nu\gamma_5u_c\right]\left[\bar{d}_d\gamma_\mu s_d\right],\\
\xi_{\mu}^{\Xi_{cc}^\ast\bar{K}}&=&\frac{1}{\sqrt{3}}\epsilon_{abc}\left[2\left(u^T_aC\gamma_\mu c_b\right)\gamma_5c_c+\left(c^T_aC\gamma_\mu c_b\right)\gamma_5u_c\right]\left[\bar{d}_d\gamma_5s_d\right],\\
\nonumber \xi^{\Xi_{cc}^\ast\bar{K}^\ast}&=&\frac{1}{\sqrt{3}}\epsilon_{abc}\left[2\left(u^T_aC\gamma_\mu c_b\right)\gamma_5c_c+\left(c^T_aC\gamma_\mu c_b\right)\gamma_5u_c\right]\left[\bar{d}_d\gamma_\mu s_d\right],\\
\nonumber \xi_{\mu\nu}^{\Xi_{cc}^\ast\bar{K}^\ast}&=&\frac{1}{\sqrt{3}}\epsilon_{abc}\left[2\left(u^T_aC\gamma_\mu c_b\right)\gamma_5c_c+\left(c^T_aC\gamma_\mu c_b\right)\gamma_5u_c\right]\left[\bar{d}_d\gamma_\nu s_d\right]+(\mu\leftrightarrow\nu),
\end{eqnarray}}
and the $\Omega_{cc}^{(\ast)}\pi/\rho$ states with interpolating currents $\psi_i$ obtained as $\psi_i=\xi_i\;(u\leftrightarrow s)$. In the above currents, 
$u, d, s, c$ denotes up, down, strange or charm quark field, respectively. $T$ denotes the transpose of a quark field, $C$ is the charge conjugation operator. The subscripts $a, b, c, d$ are the color indices. 
As fermionic operators,  these interpolating currents could couple to both the negative-parity and positive-parity pentaquark states via different coupling relations~\cite{Chung1982,Bagan1993,Jido1996,Ohtani:2012ps}
\begin{equation}\label{couple1}
    \begin{split}
      &\langle 0|J_-|X_{1/2^-}\rangle=f_{X}^-u(p)\, ,\\
         &\langle 0|J_-|X_{1/2^+}\rangle=f_{X}^+\gamma_5u(p)\, ,\\
          \end{split}
\end{equation}
in which $u(p)$ is the Dirac spinor and $f_X^\mp$ is the coupling constant. 
The two-point correlation functions induced by the above interpolating currents are~\cite{Shifman1979,Reinders1985}
\begin{equation}\label{correlation}
    \begin{split}
    \Pi(p^2)&=i\int  d^4x e^{ip\cdot x}\langle 0|T\left[J(x) \bar{J}(0)\right]|0\rangle\\
         &=\left(\hat{p}+M_{X} \right) \Pi^{1/2}(p^2)\,,
    \end{split}
    \end{equation}
\begin{equation}\label{correlation1}
    \begin{split}
    \Pi_{\mu\nu }(p^2)&=i\int d^4x e^{ip\cdot x}\langle 0|T\left[J_{\mu}(x) \bar{J}_{\nu }(0)\right]|0\rangle\\
         &=\left(\frac{p_{\mu}p_{\nu}}{p^2 }-g_{\mu\nu }\right) \left(\hat{p}+M_{X} \right) \Pi^{3/2}(p^2)+\cdots\, ,
    \end{split}
    \end{equation}
\begin{equation}\label{correlation2}
    \begin{split}
    \Pi_{\mu\nu\alpha \beta }(p^2)&=i\int d^4x e^{ip\cdot x}\langle 0|T\left[J_{\mu\nu}(x) \bar{J}_{\alpha \beta }(0)\right]|0\rangle\\
    &=\left(g_{\mu\alpha } g_{\nu\beta }+g_{\mu\beta }g_{\nu\alpha }\right) \left(\hat{p} +M_{X} \right) \Pi^{5/2}(p^2)+\cdots \, ,
    \end{split}
    \end{equation}
in which the $\Pi^{1/2}(p^2)$, $\Pi^{3/2}(p^2)$ and $\Pi^{5/2}(p^2)$ are the invariant functions for intermediate states with spin-1/2, 3/2 and 5/2 respectively. 
At the hadronic level, the correlation function can be described by the dispersion relation
\begin{equation}\label{correlationhad}
    \begin{split}
    \Pi(p^2)&=\frac{1}{\pi } \int^{\infty }_{4m_c^2} \frac{\text{Im}\Pi (s)}{s-p^2-i\varepsilon  }ds\, .
    \end{split}
    \end{equation}
The spectral function is defined as the following in the ``narrow resonance'' approximation 
\begin{equation}\label{spectral}
        \begin{split}
          \rho _{phen}(s) \equiv \frac{\text{Im}\Pi (s)}{\pi   }=\sum_{n }\delta (s-M_{n}^2) \langle 0|J|n \rangle \langle n|\bar{J}|0 \rangle  
     =f_{X}^2\delta (s-M_{X}^2)+\cdots\, .
        \end{split}
    \end{equation}

Considering both the contributions from negative-parity and positive-parity pentaquarks
\begin{equation} \label{correlationpn}
\Pi(p^2)=f_{X}^{-2}\frac{\hat p+M_X^-}{p^2-(M_X^-)^2}+f_{X}^{+2}\frac{\hat p-M_X^+}{p^2-(M_X^+)^2}+\cdots,\, 
\end{equation}
One can obtain the hadronic spectral densities in the rest frame $\overrightarrow{p}=0$~\cite{Jido1996}
\begin{equation}\label{spectralhadron}
  \begin{split}
    \frac{\text{Im}\Pi (p_0)}{\pi   }=f_{X}^{-2}\frac{\gamma _{0}p_0+M_{-}}{2}\delta (s-M_{-}^2)+f_{X}^{+2}\frac{\gamma _{0}p_0-M_{+}}{2}\delta (s-M_{+}^2)  
    = \gamma _{0}p_0  \rho _{H}^1(s) - \rho _{H}^0(s), 
    \end{split}
\end{equation}
where
\begin{equation}\label{spectral2}
  \begin{split}
    p_0\rho _{H}^1(s)=& \frac{1}{2}\left[f_{X}^{-2}\delta (s-M_{-}^2)+f_{X}^{+2}\delta (s-M_{+}^2)\right],  \\
    \rho _{H}^0(s)=& \frac{1}{2}\left[f_{X}^{-2}\delta (s-M_{-}^2)-f_{X}^{+2}\delta (s-M_{+}^2)\right].  \\
    \end{split}
\end{equation}
At the hadronic side
\begin{equation}\label{srhadron1}
  \begin{split}
    \int _{4m_{c}^2}^{s_{0}} \left[\sqrt{s} \rho _{j,H}^1(s)+\rho _{j,H}^0(s) \right] \text{exp}\left(-\frac{s}{M_B^2}\right)ds=2M_{-}f_{j,X}^{-2}\text{exp}\left(-\frac{M_{-}^2}{M_B^2}\right) \, ,
     \end{split}
\end{equation}
\begin{equation}\label{srhadron2}
  \begin{split}
    \int _{4m_{c}^2}^{s_{0}} \left[\sqrt{s} \rho _{j,H}^1(s)-\rho _{j,H}^0(s) \right] \text{exp}\left(-\frac{s}{M_B^2}\right)ds=2M_{+}f_{j,X}^{+2}\text{exp}\left(-\frac{M_{+}^2}{M_B^2}\right) \, ,
     \end{split}
\end{equation}
for the negative-parity and positive-parity state respectively with spin $j=1/2,\, 3/2,\, 5/2$.  

At the quark-gluonic side, we calculate spectral functions via the operator product expansion (OPE) method up to the dimension 10 condensates.
Using the quark-hadron duality, one can establish the QCD sum rules for hadron mass 
\begin{equation}\label{massQCD}
  \begin{split}
     M_{j, \pm }^2=\frac{    \int _{4m_{c}^2}^{s_{0}} \left[\sqrt{s} \rho _{j,QCD}^1(s)\mp \rho _{j,QCD}^0(s) \right] \text{exp}\left(-\frac{s}{M_B^2}\right)sds}
     {\int _{4m_{c}^2}^{s_{0}} \left[\sqrt{s} \rho _{j,QCD}^1(s)\mp \rho _{j,QCD}^0(s) \right] \text{exp}\left(-\frac{s}{M_B^2}\right)ds}\, ,
        \end{split}
\end{equation}
in which $s_0$ and $M_B$ are the continuum threshold and Borel mass, respectively.
  
The $\Sigma _{c}^{(*)}D^{(*)}$ pentaquarks can be isospin quartet $(P_{cc}^{+++}, \, P_{cc}^{++}, \, P_{cc}^{+}, \, P_{cc}^{0})$ with $I=3/2$ or doublet $(P_{cc}^{++}, \, P_{cc}^{+})$ with $I=1/2$. The pentaquarks in the isospin quartet do not mix with the ordinary doubly charmed baryons, since they have different flavor quantum numbers. The triply charged $P_{cc}^{+++}(ccuu\bar d)$ and neutral $P_{cc}^{0}(ccdd\bar u)$ states can be considered as the characteristic signals for these doubly charmed pentaquarks. 

\section{Numerical results}
We use the following parameter values to perform numerical analyses $\langle \bar{q}q\rangle(1\mathrm{GeV})=-(0.24\pm0.03)^3\;\mathrm{GeV}^3$, $\langle \bar{q}g_s\sigma\cdot Gq\rangle(1\mathrm{GeV})=-M_0^2\langle \bar{q}q\rangle$, $M_0^2=(0.8\pm0.2)\;\mathrm{GeV}^2$, $\langle \bar{s}s\rangle/\langle \bar{q}q\rangle=0.8\pm0.1$, $\langle g_s^2GG\rangle(1\mathrm{GeV})=(0.48\pm0.14)\;\mathrm{GeV}^4$ and $m_s(2\;\mathrm{GeV})=95^{+9}_{-3}\;\mathrm{MeV}$, $m_c(m_c)=1.27^{+0.03}_{-0.04}\;\mathrm{GeV}$, $m_b(m_b)=4.18_{-0.03}^{+0.04}\;\mathrm{GeV}$~\cite{PDG,Shifman1979,Reinders1985}. To ensure the good behaviors of mass sum rules, we investigate the OPE convergence, pole contribution and stability of Borel curves to determine the working regions of the continuum threshold $s_0$ and Borel parameter $M_B^2$. 

\begin{figure}[t!]
  \centering
  \includegraphics[width=6.5cm]{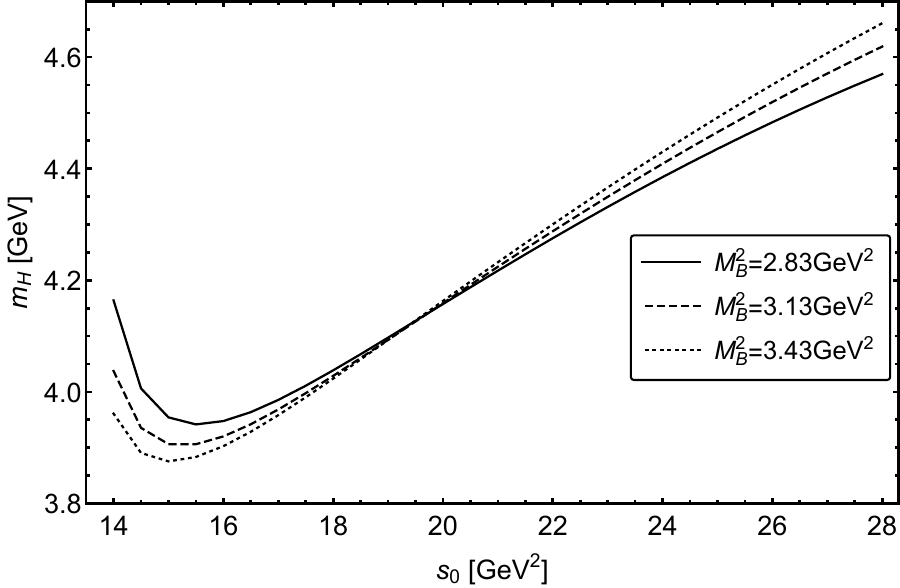}\quad
  \includegraphics[width=6.5cm]{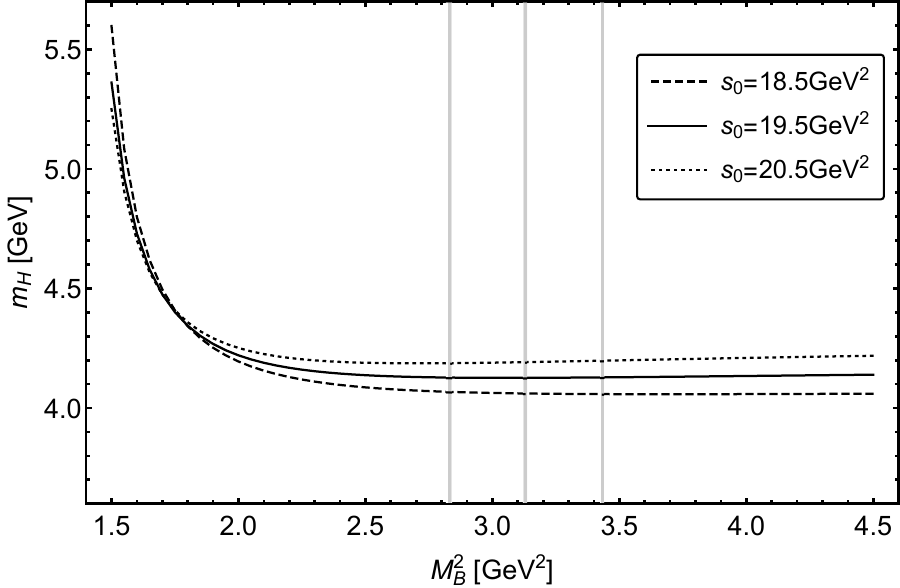}\\
  \caption{Variations of hadron mass to $s_0$ and $M_B^2$ for $J^{\Lambda _cD}$ with
  $J^P=\frac{1}{2}^-$.
    }
  \label{fig:lanmdc-D-mass-s0}
  \end{figure}

As an example, we show the variations of hadron mass to $s_0$ and $M_B^2$ for the interpolating current $J^{\Lambda _cD}$ with $J^P=\frac{1}{2}^-$ in Fig.~\ref{fig:lanmdc-D-mass-s0}. One finds that the Borel curves stability can be well ensured in the Borel window to give reliable mass prediction. After perform numerical analyses, we obtain the pentaquark masses for all interpolating currents in Table~\ref{PccfumassTab}, Table~\ref{PcczhmassTab} and Table~\ref{Tab:Result1-}, along with corresponding two-hadron mass thresholds in the last columns for various channels. It is shown that the predicted masses for the $\Lambda _cD\, (\frac{1}{2}^-)$, $\Sigma _cD\, (\frac{1}{2}^-)$, $\Sigma _cD^*\, (\frac{3}{2}^-)$, $\Lambda _c^*D\, (\frac{3}{2}^-)$, $\Lambda _c^*D^*\, (\frac{5}{2}^-)$ and $\Sigma _cD\, (\frac{1}{2}^+)$, $\Sigma _cD^\ast\, (\frac{3}{2}^+)$, $\Sigma _c^\ast D\, (\frac{3}{2}^+)$ molecular pentaquarks without strangeness and $\Xi_cD^{\ast}\, (\frac{1}{2}^-)$, $\Xi_c^{'}D^{\ast}\, (\frac{3}{2}^-)$, $\Xi_{c}^{\ast}D^{\ast}\, (\frac{1}{2}^-)$, $\Xi_{cc}^{\ast}\bar{K}^{\ast }\, (\frac{1}{2}^-, \frac{5}{2}^-)$, $\Omega_{cc}^{\ast }\rho\, (\frac{1}{2}^-, \frac{5}{2}^-)$ molecular pentaquarks with strangeness $S=-1$ are lower than their meson-baryon mass thresholds, implying the existence of these doubly charmed $P_{cc}$ and $P_{ccs}$ bound states in these channels. 

 \begin{table}[h!]
      \caption{Predicted masses for the $\Lambda _{c}^{(*)}D^{(*)}$ and $\Sigma _{c}^{(*)}D^{(*)}$ molecular pentaquarks with negative-parity.}\label{PccfumassTab}
      \renewcommand\arraystretch{1.3} 
      \setlength{\tabcolsep}{0.4 em}{ 
      \begin{center}
      \begin{tabular}{c c c c c c}
        \hline
         \hline
       current& $J^{P}$ & $s_0 [\mathrm{GeV}^2]$ & $M_B^2 [\mathrm{GeV}^2]$ &  Mass $[\mathrm{GeV}]$ & Threshold $[\mathrm{GeV}]$   \\
        \hline
       $J^{\Lambda _{c}D}$    & $\frac{1}{2}^-$ & $19.5 $ & 2.83$\sim $3.43  & $4.13_{-0.09}^{+0.10}$ & 4.15 \\
       $J^{\varSigma _{c}D}$    & $\frac{1}{2}^-$ & $18.3 $ & 3.40$\sim $3.70  & $4.08_{-0.13}^{+0.18}$ & 4.32 \\
       $J^{\varSigma _{c}D^*}$    & $\frac{3}{2}^-$ & $20.3 $ & 3.17$\sim $3.47  & $4.14_{-0.15}^{+0.18}$ & 4.46 \\
       $J^{\varSigma _{c}^*D}$    & $\frac{3}{2}^-$ & $22.8 $ & 3.82$\sim $4.22  & $4.47_{-0.10}^{+0.11}$ & 4.39\\
       $J^{\Lambda _{c}D^*}$   & $\frac{3}{2}^-$ & $21.0 $ & 3.55$\sim $3.95  & $4.31_{-0.10}^{+0.11}$ & 4.29 \\
       $J^{\Lambda _{c}^*D}$  & $\frac{3}{2}^-$ & $22.8(\pm 5\%)$ & 2.91$\sim $3.51  & $4.42_{-0.12}^{+0.13}$ & 4.73\\  
       $J^{\Lambda _{c}^*D^*}$  & $\frac{5}{2}^-$ & $22.1 $ & 3.09$\sim $3.69  & $4.41_{-0.14}^{+0.17}$ & 4.86 \\   
       $J^{\varSigma _{c}^*D^*}$    & $\frac{5}{2}^-$ & $25.0 $ & 4.0$\sim $4.6  & $4.69_{-0.11}^{+0.12}$ & 4.53\\
       \hline
         \hline
      \end{tabular}
      \end{center}}
      \end{table}

\begin{table}[h!]
        \caption{Predicted masses for the $\Lambda _{c}^{(*)}D^{(*)}$ and $\Sigma _{c}^{(*)}D^{(*)}$ molecular pentaquarks with positive-parity.}\label{PcczhmassTab}
        \renewcommand\arraystretch{1.3} 
        \setlength{\tabcolsep}{0.4 em}{ 
        \begin{center}
        \begin{tabular}{c c c c c c}
          \hline
           \hline
         current& $J^{P}$ & $s_0 [\mathrm{GeV}^2]$ & $M_B^2 [\mathrm{GeV}^2]$ &   Mass $[\mathrm{GeV}]$ & Threshold $[\mathrm{GeV}]$ \\
          \hline
         $J^{\Lambda _{c}D}$    & $\frac{1}{2}^+$ & $27.4 $ & 4.11$\sim $4.51  & $4.97_{-0.14}^{+0.13}$ & 4.46  \\
         $J^{\varSigma _{c}D}$    & $\frac{1}{2}^+$ & $23.7 $ & 2.98$\sim $3.58  & $4.60_{-0.12}^{+0.11}$ & 4.63 \\
         $J^{\varSigma _{c}D^*}$    & $\frac{3}{2}^+$ & $25.6 $ & 4.00$\sim $4.30  & $4.77_{-0.09}^{+0.09}$ & 4.77 \\
         $J^{\varSigma _{c}^*D}$    & $\frac{3}{2}^+$ & $24.9 $ & 3.61$\sim $4.01  & $4.64_{-0.21}^{+0.17}$ & 4.64 \\
         $J^{\Lambda _{c}D^*}$   & $\frac{3}{2}^+$ & $35.0 $ & 4.24$\sim $4.84  & $5.52_{-0.11}^{+0.11}$ & 4.60\\
         $J^{\Lambda _{c}^*D}$  & $\frac{3}{2}^+$ & $22.9 $ & 3.06$\sim $3.56  & $4.54_{-0.18}^{+0.15}$ & 4.50\\  
         $J^{\Lambda _{c}^*D^*}$  & $\frac{5}{2}^+$ & $33.0 $ & 2.99$\sim $3.59  & $5.29 _{-0.14}^{+0.14}$ & 4.64 \\   
         $J^{\varSigma _{c}^*D^*}$    & $\frac{5}{2}^+$ & $28.8 $ & 3.36$\sim $3.96  & $5.06_{-0.30}^{+0.24}$ & 4.78 \\
         \hline
           \hline
        \end{tabular}
        \end{center}}
        \end{table}
\begin{table}[h!]
\caption{Predicted masses for the $\Xi_{c}^{(\prime\ast)}D^{(\ast)}$ and $\Omega_{cc}^{(\ast)}\pi/\rho$ molecular pentaquarks with negative-parity.}\label{Tab:Result1-}\renewcommand\arraystretch{1.3} 
\setlength{\tabcolsep}{0.4 em}{ 
\begin{center}
\begin{tabular}{c c c c c c}
  \hline
 \hline
Current & $J^P$ & $M_B^2[\mathrm{GeV}^2]$ & $s_0[\mathrm{GeV}^2]$ & Mass[GeV] & Threshold[MeV] \\ 
 \hline
 $\eta^{\Xi_{c} D^{\ast}}$ & $\frac{1}{2}^-$ & 4.31 & 22.3 & $4.38^{+0.05}_{-0.04}$ & 4477\\
 $\eta_{\mu}^{\Xi_{c}^\prime D^{\ast}}$ & $\frac{3}{2}^-$ & 4.16 & 22.3 & $4.43^{+0.05}_{-0.04}$ & 4588\\
 $\eta^{\Xi_{c}^\ast D^{\ast}}$ & $\frac{1}{2}^-$ & 3.58 & 24.5 & $4.56^{+0.06}_{-0.05}$ & 4655\\
 $\xi^{\Xi_{cc} \bar{K}}$ & $\frac{1}{2}^-$ & 4.19 & 20.3 & $4.20^{+0.05}_{-0.05}$ & 4120\\
 $\xi_{\mu}^{\Xi_{cc} \bar{K}^\ast}$ & $\frac{3}{2}^-$ & 3.89 & 24.3 & $4.52^{+0.06}_{-0.05}$ & 4512\\
 $\xi_{\mu}^{\Xi_{cc}^\ast \bar{K}}$ & $\frac{3}{2}^-$ & 4.13 & 21.3 & $4.28^{+0.05}_{-0.05}$ & 4192\\
 $\xi^{\Xi_{cc}^\ast \bar{K}^\ast}$ & $\frac{1}{2}^-$ & 3.56 & 22.3 & $4.37^{+0.05}_{-0.05}$ & 4584\\
 $\xi_{\mu\nu}^{\Xi_{cc}^\ast \bar{K}^\ast}$ & $\frac{5}{2}^-$ & 4.16 & 22.3 & $4.34^{+0.05}_{-0.04}$ & 4584\\
 $\psi^{\Omega_{cc} \pi}$ & $\frac{1}{2}^-$ & 4.16 & 21.3 & $4.27^{+0.05}_{-0.05}$ & 3853\\
 $\psi_{\mu}^{\Omega_{cc} \rho}$ & $\frac{3}{2}^-$ & 3.69 & 24.3 & $4.50^{+0.06}_{-0.06}$ & 4488\\
 $\psi_{\mu}^{\Omega_{cc}^\ast \pi}$  & $\frac{3}{2}^-$ & 4.09 & 22.3 & $4.36^{+0.05}_{-0.05}$ & 3925\\
 $\psi^{\Omega_{cc}^\ast \rho}$  & $\frac{1}{2}^-$ & 3.58 & 22.3 & $4.37^{+0.05}_{-0.05}$ & 4560\\ 
 $\psi_{\mu\nu}^{\Omega_{cc}^\ast \rho}$  & $\frac{5}{2}^-$ & 4.38 & 22.3 & $4.36^{+0.05}_{-0.05}$ & 4560\\ 
  \hline
  \hline
\end{tabular}
\end{center}}
\end{table}

\section{Summary}
Motivated by the LHCb's observations of hidden-charm pentaquark states and doubly charmed tetraquark state, we have systematically studied the mass spectra of the doubly charmed pentaquark states with strangeness $S=0, -1$. We construct the $\Lambda _{c}^{(*)}D^{(*)}$, $\Sigma _{c}^{(*)}D^{(*)}$, $\Xi_{c}^{(\prime\ast)}D^{(\ast)}$,  $\Xi_{cc}^{(\ast)}\bar{K}^{(\ast)}$ and $\Omega_{cc}^{\ast}\pi/\rho$ molecular pentaquark interpolating currents with spin-parity $J^P=1/2^\pm, 3/2^\pm, 5/2^\pm$, and calculate the two-point correlation functions and spectral functions up to dimension ten condensates. 

Our results show that the doubly charmed $\Lambda _cD\, (\frac{1}{2}^-)$, $\Sigma _cD\, (\frac{1}{2}^-)$, $\Sigma _cD^*\, (\frac{3}{2}^-)$, $\Lambda _c^*D\, (\frac{3}{2}^-)$, $\Lambda _c^*D^*\, (\frac{5}{2}^-)$, $\Sigma _cD\, (\frac{1}{2}^+)$, $\Sigma _cD^\ast\, (\frac{3}{2}^+)$, $\Sigma _c^\ast D\, (\frac{3}{2}^+)$ and $\Xi_cD^{\ast}\, (\frac{1}{2}^-)$, $\Xi_c^{'}D^{\ast}\, (\frac{3}{2}^-)$, $\Xi_{c}^{\ast}D^{\ast}\, (\frac{1}{2}^-)$, $\Xi_{cc}^{\ast}\bar{K}^{\ast }\, (\frac{1}{2}^-, \frac{5}{2}^-)$, $\Omega_{cc}^{\ast }\rho\, (\frac{1}{2}^-, \frac{5}{2}^-)$ pentaquarks are predicted to lie below the corresponding meson-baryon mass thresholds, implying the possibility of bound states in these channels. Especially for the  triply charged $P_{cc}^{+++}(ccuu\bar d)$ and neutral $P_{cc}^{0}(ccdd\bar u)$ states in the isospin quartet, they do not mix with the ordinary doubly charmed baryons $\Xi_{cc}^{++}/\Xi_{cc}^{+}$ due to their exotic flavors and charges. We suggest search for these characteristic pentaquark signals in the $P_{cc}^{+++}\to\Xi_{cc}^{(\ast) ++}\pi^+/\rho^+$, $\Sigma_c^{(\ast)++}D^{(\ast)+}$ and $P_{cc}^{0}\to\Xi_{cc}^{(\ast) +}\pi^-/\rho^-$, $\Sigma_c^{(\ast)0}D^{(\ast)0}$ decay processes. 

\section*{ACKNOWLEDGMENTS}
This work is supported in part by the National Natural Science Foundation of China with Grant Nos. 12175318, 12375073 and 12035007, Guangdong Provincial funding with Grant Nos. 2019QN01X172, Guangdong Major Project of Basic and Applied Basic Research No. 2020B0301030008, the Natural Science Foundation of Guangdong Province of China under Grant No. 2022A1515011922, the NSFC and the Deutsche Forschungsgemeinschaft (DFG, German
Research Foundation) through the funds provided to the Sino-German Collaborative
Research Center TRR110 ``Symmetries and the Emergence of Structure in QCD''
(NSFC Grant No. 12070131001, DFG Project-ID 196253076-TRR 110).

\bibliography{MyRef.bib}
\bibliographystyle{JHEP}

%

\end{document}